\documentclass[conference]{IEEEtran}
\IEEEoverridecommandlockouts

\usepackage{cite}
\usepackage{amsmath,amssymb,amsfonts}
\usepackage{algorithmic}
\usepackage{graphicx}
\usepackage{textcomp}
\usepackage[table]{xcolor}     
\usepackage{float}
\usepackage{colortbl}          
\usepackage{array}

\def\BibTeX{{\rm B\kern-.05em{\sc i\kern-.025em b}\kern-.08em
    T\kern-.1667em\lower.7ex\hbox{E}\kern-.125emX}}

 \usepackage[letterpaper, top=0.77in, bottom=1.05in, left=0.75in, right=0.75in]{geometry}

\begin{document}

\title{Graph Based Deep Reinforcement Learning Aided by Transformers for Multi-Agent Cooperation\\
\thanks{This material is based upon the work supported by the National Science Foundation under Grant Number 2204721 and MIT Lincoln Laboratory under Grant Number 7000612889.}
}

\author{
    \IEEEauthorblockN{
        Michael Elrod\IEEEauthorrefmark{1},
        Niloufar Mehrabi\IEEEauthorrefmark{1},
        Rahul Amin\IEEEauthorrefmark{3},  
        Manveen Kaur\IEEEauthorrefmark{2},
        Long Cheng\IEEEauthorrefmark{1},
        Jim Martin\IEEEauthorrefmark{1},
        Abolfazl Razi\IEEEauthorrefmark{1}
    }
    \IEEEauthorblockA{
        \IEEEauthorrefmark{1}School of Computing, Clemson University, Clemson, SC, USA 
    }
    \IEEEauthorblockA{
        \IEEEauthorrefmark{2}Computer Science Department, California State University, Los Angeles, CA, USA 
    }
    \IEEEauthorblockA{
        \IEEEauthorrefmark{3}Lincoln Laboratory, Massachusetts Institute of Technology, Lexington, MA, USA 
    }
}

\maketitle

\begin{abstract}
Mission planning for a fleet of cooperative autonomous drones in applications that involve serving distributed target points, such as disaster response, environmental monitoring, and surveillance, is challenging, especially under partial observability, limited communication range, and uncertain environments. Traditional path-planning algorithms struggle in these scenarios, particularly when prior information is not available. To address these challenges, we propose a novel framework that integrates Graph Neural Networks (GNNs), Deep Reinforcement Learning (DRL), and transformer-based mechanisms for enhanced multi-agent coordination and collective task execution. Our approach leverages GNNs to model agent-agent and agent-goal interactions through adaptive graph construction, enabling efficient information aggregation and decision-making under constrained communication. A transformer-based message-passing mechanism, augmented with edge-feature-enhanced attention, captures complex interaction patterns, while a Double Deep Q-Network (Double DQN) with prioritized experience replay optimizes agent policies in partially observable environments. This integration is carefully designed to address specific requirements of multi-agent navigation, such as scalability, adaptability, and efficient task execution. Experimental results demonstrate superior performance, with 90\% service provisioning and 100\% grid coverage (node discovery), while reducing the average steps per episode to 200, compared to 600 for benchmark methods such as particle swarm optimization (PSO), greedy algorithms and DQN.

\end{abstract}

\begin{IEEEkeywords}
Graph Neural Networks (GNNs), Adaptive Multi-Agent Navigation, Transformer-Enhanced Path Planning, Collaborative Decision-Making
\end{IEEEkeywords}

\section{Introduction}
Autonomous multi-drone systems are increasingly critical for large-scale applications such as disaster response, environmental monitoring, and precision agriculture \cite{boroujeni2024comprehensive, boroujeni2025fire}. These systems enable collaborative task execution—such as crop pollination, ecosystem restoration, and urban farming support—by efficiently navigating distributed nodes across expansive geographic areas. However, their operational efficacy is constrained by two interdependent challenges: limited onboard energy resources (which also implies limited communication ranges) and partial observability (dictated by flight altitude and camera field of view). The latter is imperative when serving areas are not determined in advance.

While multi-agent coordination frameworks often assume seamless information exchange, real-world scenarios necessitate using lightweight communication protocols. Bandwidth limitations prevent the transmission of high-resolution imagery, requiring drones to share only compact representations such as trajectories, task status updates, or model parameters. This partial observability complicates coordination, demanding robust decision-making policies that balance energy efficiency, adaptability to dynamic environments, and mission-critical reliability.

Path planning is often a core component of mission planning for autonomous systems, alongside other aspects such as resource allocation, task scheduling, and coordination. Conventional methods for path planning, such as Traveling Salesman Problem (TSP) solvers \cite{pop2024comprehensive}, numerical optimization \cite{qin2023review}, heuristic approaches \cite{abdulsaheb2023classical}, and greedy algorithms \cite{xiang2022combined}, have been widely used but often struggle with dynamic and probabilistic environments, especially when service points are not determined prior. In recent years, Deep Reinforcement Learning (DRL) has gained prominence as a powerful alternative, demonstrating the ability to handle probabilistic situations, develop generalizable policies, and learn from experience under partial observability \cite{mehrabi2024adaptive}. Its ability to integrate with visual input and sensor data further enhances its suitability for complex mission planning in dynamic settings. 
However, DRL techniques often demand substantial training time and computational resources, posing challenges for real-world deployment. Moreover, in multi-agent systems, DRL can face difficulties with coordination and generalization, especially in environments characterized by partial observability or restricted communication. These limitations highlight the need for more advanced algorithms to exploit long-term cooperation benefits, an area where transformers have proven to flourish.

Several prior studies have addressed multi-drone path planning and coordination, but significant gaps remain. Centralized approaches \cite{munoz2021multi} efficiently manage trajectories, yet they suffer from scalability issues and single points of failure. In contrast, decentralized methods \cite{azam2021decentralized} enhance robustness by enabling Unmanned Aerial Vehicles (UAVs) to operate autonomously and make decisions based on local observations, but full information exchange is often impractical. Moreover, limited coordination among UAVs can lead to inefficiencies and suboptimal coverage. Although Decentralized Markov Decision Processes (Dec-MDPs) \cite{azam2021decentralized} show promise for multi-goal tracking, they still struggle with collaborative decision-making and minimizing path redundancy. Furthermore, communication-aware methods \cite{s23052560}, such as Exact Dubins Multi-Robot Coverage Path Planning (EDM) and Credit-based Dubins Multi-Robot Coverage Path Planning (CDM) algorithms, partition regions and balance task allocation using heuristic and exact approaches. While these methods effectively reduce computation and improve task balance, they lack adaptability to dynamic environments and address real-time coordination challenges in heterogeneous systems.

Similarly, reinforcement learning-based strategies, such as those proposed in \cite{zhang2013coordinating}, demonstrate promise in autonomous decision-making by enabling agents to dynamically identify coordination sets and optimize joint policies in cooperative environments. However, these approaches often struggle with scalability in large multi-agent systems and encounter challenges in balancing communication costs with learning performance. For instance, using Distributed Constraint Optimization Problems (DCOP) \cite{zhang2013coordinating} techniques can improve coordination but may lead to significant communication overhead, making real-time applications infeasible under limited bandwidth conditions. 
These limitations highlight the need for more elegantly designed frameworks, leveraging computational intelligence and lightweight information exchange among drones to exploit long-term cooperation benefits via robust and coordinated trajectory optimization.

To address these challenges, 
this paper presents a novel framework that integrates Graph Neural Networks (GNNs) with a Transformer-enhanced Deep Q-learning (DQN) approach to harmonically exploit benefits from both spatial and temporal relations among drone trajectories. The proposed method leverages the strengths of these components to enable efficient information exchange, coordination, and decision-making among agents. 

The utilized GNN operates on a graph-structured representation of the environment, where agents and goals are modeled as nodes, and their relationships are encoded as weighted edges. Using a message-passing mechanism, the GNN iteratively updates node embeddings by aggregating local observations and semantic information from neighboring nodes. This iterative process enables agents to exchange lightweight, high-level information, enhancing their understanding of both their immediate surroundings and the broader environment. Using attention mechanisms within the GNN, drones prioritize a subset of high-value goals within their visual range, optimizing their focus to achieve better overall task performance. 
The output of the GNN forms the input sequence for a transformer-based architecture, which processes the embeddings to capture long-range dependencies and optimize sequential decision-making. By modeling interactions across the entire graph, the transformer complements the GNN’s local information exchange, refining the agents’ policies to account for global relationships and long-term strategies. 
Finally, the refined embeddings are fed into a Deep Q-network, which computes Q-values for action selection, enabling agents to optimize their trajectories and behaviors. This integration of GNNs and transformers with DQN enhances the framework's scalability, coordination, and adaptability, making it particularly well-suited for dynamic, partially observable, and resource-constrained environments.

The main contributions of this work include: 
\begin{itemize}
    \item We propose a novel graph neural network architecture combining an entity-specific embedding layer and a transformer-based message-passing mechanism to enable efficient inter-agent communication and decision-making in dynamic environments.
    \item We introduce an adaptive graph construction method based on distance thresholding and nearest-neighbor constraints. This ensures efficient representation of agent-agent and agent-goal interactions while accommodating the drones' limited communication range.
    \item We leverage multi-head transformer layers with edge-feature-enhanced attention mechanisms to effectively capture complex interactions in the graph structure.
    \item We utilize Double DQN with prioritized experience replay to optimize agent policies, ensuring robust and efficient learning in partially observable environments.

\end{itemize}


\section{System Model}

The proposed framework operates within a grid-based environment, implemented using the MiniGrid library in Python. 
The environment consists of an $L \times L$ grid, where $L$ represents the dimensions of the square world. Within this environment, $N$ autonomous agents must cooperatively discover and visit (serve) the location of $M$ goals distributed randomly throughout the grid. Note that nodes are discovered when fall within a drone's visual range but are served only when a drone enters their hosting cell.

\subsection{Environment Structure}
Each cell in the $L \times L$ grid can be occupied by at most one agent or goal at any given time. At the start of each episode, agents are distributed uniformly over the four borderlines of the environment, to mimic realistic situations (e.g., disaster management) where typically drone stations are outside the coverage area. 
Goal locations are uniformly distributed within the interior of the confined grid at the start of each episode. When a goal is visited by any agent, it is marked as served (or collected) but remains visible for the remainder of the episode.

\subsection{Agent Capabilities and Constraints}
To operate effectively in a partially observable environment, each agent is subject to two key constraints, including i) vision range: a circular field of view with radius $r_v$ centered on their current position, and ii) communication constraint: a limit of $k$ nearest neighbor drones that an agent can communicate with, where $k$ represents the maximum number of outgoing communication links per agent. Goals within the visual range of any drone are marked "visited", and are included in the subsequent decision-making by the GNN-based framework. Each drone is allowed to communicate with only $k$-nearest neighbor drones noting the limited communication bandwidth, thereby forming a dynamic and directed contact graph. 

\begin{figure}[!h]
\centering
\includegraphics[width=0.68\columnwidth]{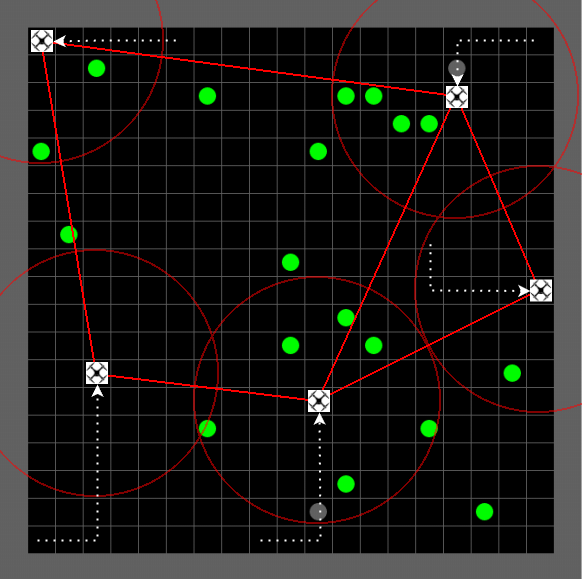}
\caption{Example configuration of the grid environment with 5 agents and 20 goals, uncollected goals (green), collected goals (grey), and agent vision radius $r_v$ (red circle around agents). Drones are allowed to communicate with at most $k=3$ adjacent drones.}
\label{fig_env}
\end{figure}

\section{Methodology}

\subsection{Problem Formulation}

The multi-agent navigation problem follows Markov Decision Process (MDP) formulation with partial observability, where each agent's observation is limited to its local field of view. The MDP is defined by the tuple $(S, A, T, R)$, where $S$ represents the state space, $A$ defines the action space, $T$ denotes the transition dynamics, and $R$ specifies the reward function.
The state space, $S$ is defined as the set of all possible states in the environment. For each agent $i$ at time $t$, its state is represented as:
\begin{equation}
    s_i(t) = (p_i(t), G_i(t), V_i(t)) \in S
\end{equation}

where $p_i(t) = (x_i(t), y_i(t))$ denotes agent $i$'s position, $G_i(t)$ represents the set of visible and uncollected goals, and $V_i(t)=\{(v,w): \sqrt{(v-x_i(t))^2+(w-y_i(t))^2} \leq r_v\}$ defines the agent's visible region. For notational simplicity, we may omit the time parameter $(t)$ when it is clear from the context. The visibility is limited to a radius of $r_v = 4.5$ units, ensuring partial observability of the environment.

The action space, $A$, consists of four discrete actions: up, down, left, and right.
The reward function $R$ is designed to encourage goal-directed behavior while penalizing invalid actions. Specifically, the 
The reward $R$ includes the accumulation of $\gamma^t(r_g I_g + r_v I_v)$ terms, 
where $\gamma = 0.99$ is the discount factor, and $t$ is the current time step. The term $r_g = +10$ provides a reward for successfully collecting a goal, while $r_v = -5$ applies a penalty for invalid moves. The indicator functions $I_g$ and $I_v$ are used to determine whether a goal has been collected or an invalid move has been made, respectively.

\subsection{Graph Neural Network Architecture}

At the core of our method is a Deep Q-Network (DQN), where a Graph Neural Network (GNN) replaces the traditional CNN as the feature extractor. The GNN processes graph-structured data, embedding agent-to-agent and agent-to-goal relationships into representative semantic vectors. These embeddings serve as the foundation for decision-making in the DQN, enabling it to compute Q-values and optimize policies with enhanced scalability and performance in multi-agent environments. The architecture comprises two key components: (i) an embedding layer to process entity-specific features and (ii) a transformer-based message-passing mechanism to capture long-range dependencies. The network operates on a dynamic graph $G = (V, E)$, where $V$ represents nodes (agents and goals) and $E$ represents weighted edges (relationships).

\begin{figure}[!h]
\centering
\includegraphics[width=0.9\columnwidth]{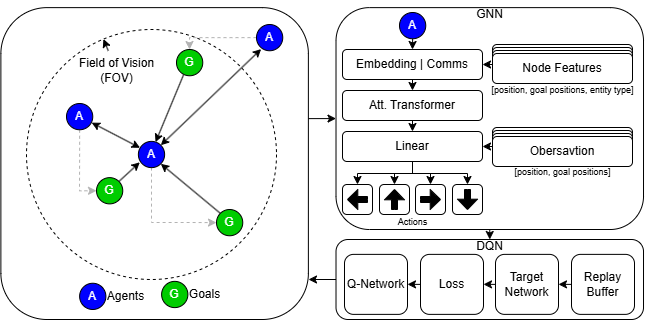}
\caption{Proposed GNN architecture consisting of an entity embedding layer and transformer-based message passing mechanism.}
\label{fig_gnn}
\end{figure}

As shown in Fig.~\ref{fig_gnn}, the GNN iteratively updates node embeddings through message passing, aggregating local observations, and compiling them into semantic information by sharing them with neighboring nodes. These updated embeddings are passed to the transformer, which processes them as a sequence to capture global dependencies. By combining localized interactions with the global context, the framework ensures efficient coordination and policy optimization in dynamic environments.

\subsection{Node Feature Construction}
For each node $v_i \in V$, a feature vector $f_i$ is constructed to capture both spatial and temporal characteristics. This vector is defined as:
\begin{equation}
    f_i = [\Delta p_i, \{g_k, s_k\}_{k=1}^3, \tau_i],
\end{equation}
where $\Delta p_i = (x_i - x_{ref}, y_i - y_{ref})$ represents the relative position of node (goal) $i$ to the observing agent at position $(x_{ref},y_{ref})$. The terms $\{g_k\}_{k=1}^3$ and $\{s_k\}_{k=1}^3$ denote the positions of the three closest visible goals and their respective collections statuses. Furthermore, $\tau_i \in \{0,1\}$ identifies whether a node represents an agent ($\tau_i=0$) or a goal ($\tau_i=1$).
This structured representation allows the model to effectively process the spatial relationships and dynamic goal-related information essential for solving the multi-agent navigation problem.

\subsection{Adaptive Graph Construction}
The graph structure is dynamically updated at each time step using a novel distance-based thresholding mechanism. Edge weights $e_{ij}$ are computed as:
\begin{equation}
    e_{ij} = \begin{cases}
        d_{ij} & \text{if } (i,j) \in E_{valid} \\
        0 & \text{otherwise}
    \end{cases}
\end{equation}

Here, $E_{valid}$ is the set of valid edges satisfying the following constraints:

\begin{enumerate}
    \item Agent-to-agent connections: $j \in \mathcal{N}_k(i)$, where $\mathcal{N}_k(i)$ is the set of $k=3$ nearest agents to agent $i$.
    \item Distance threshold: $d_{ij} \leq r_v$, where $r_v=4.5$ units
    \item Non-self-connectivity: $i \neq j$
\end{enumerate}

\subsection{Entity Embedding and Message Passing}
The network processes information hierarchically, beginning with a specialized embedding layer and followed by multi-head transformer layers to capture both local and global dependencies.

1) Entity Embedding Layer: The initial feature vector $f_i$ for each node is extended with additional context, including entity type embeddings and edge weights:
\begin{equation}
    \hat{f}_i = [f_i, \phi(\tau_i), e_{ij}],
\end{equation}
where $\phi(\tau_i)$ is a learned embedding for the entity type $\tau_i$. The term $e_{ij}$ denotes the edge weight between nodes $i$ and $j$.



2) Multi-Head Transformer Layers:
The message-passing mechanism is implemented using multi-head transformer layers. The node embeddings are updated as
\begin{equation}
    z_i^{l+1} = \sum_{h=1}^H \sigma(\sum_{j \in \mathcal{N}(i)} \alpha_{ij}^h W_h^l z_j^l),
\end{equation}
where $H=3$ is the number of attention heads and $W_i^{l}$ is the learnable weight matrix for head $h$ at layer $l$. The attention coefficients $\alpha_{ij}^h$ are computed as
\begin{equation}
    \alpha_{ij}^h = \text{softmax}_j(\frac{(W_Q^h z_i^l)(W_K^h z_j^l)^T}{\sqrt{d_k}} + b_{ij}),
\end{equation}
where $W_Q^{h}$ and $W_K^{h}$ are the query and key projection matrices for head $h$, $d_k$ is the dimensionality of the key vectors, and $b_{ij} = W_e e_{ij}$ incorporates edge features into the attention mechanism. 

In this work, we leverage attention mechanisms within a transformer-based message-passing framework to prioritize critical interactions in the graph structure. The attention weights dynamically adjust to emphasize relationships such as agent-to-goal and agent-to-agent connections based on their relevance to the decision-making process. For instance, an agent closer to a goal or with nearby collaborators receives higher attention, enabling efficient and coordinated navigation. 

\begin{figure}[htbp]
    \centering
    \includegraphics[width=0.95\columnwidth]{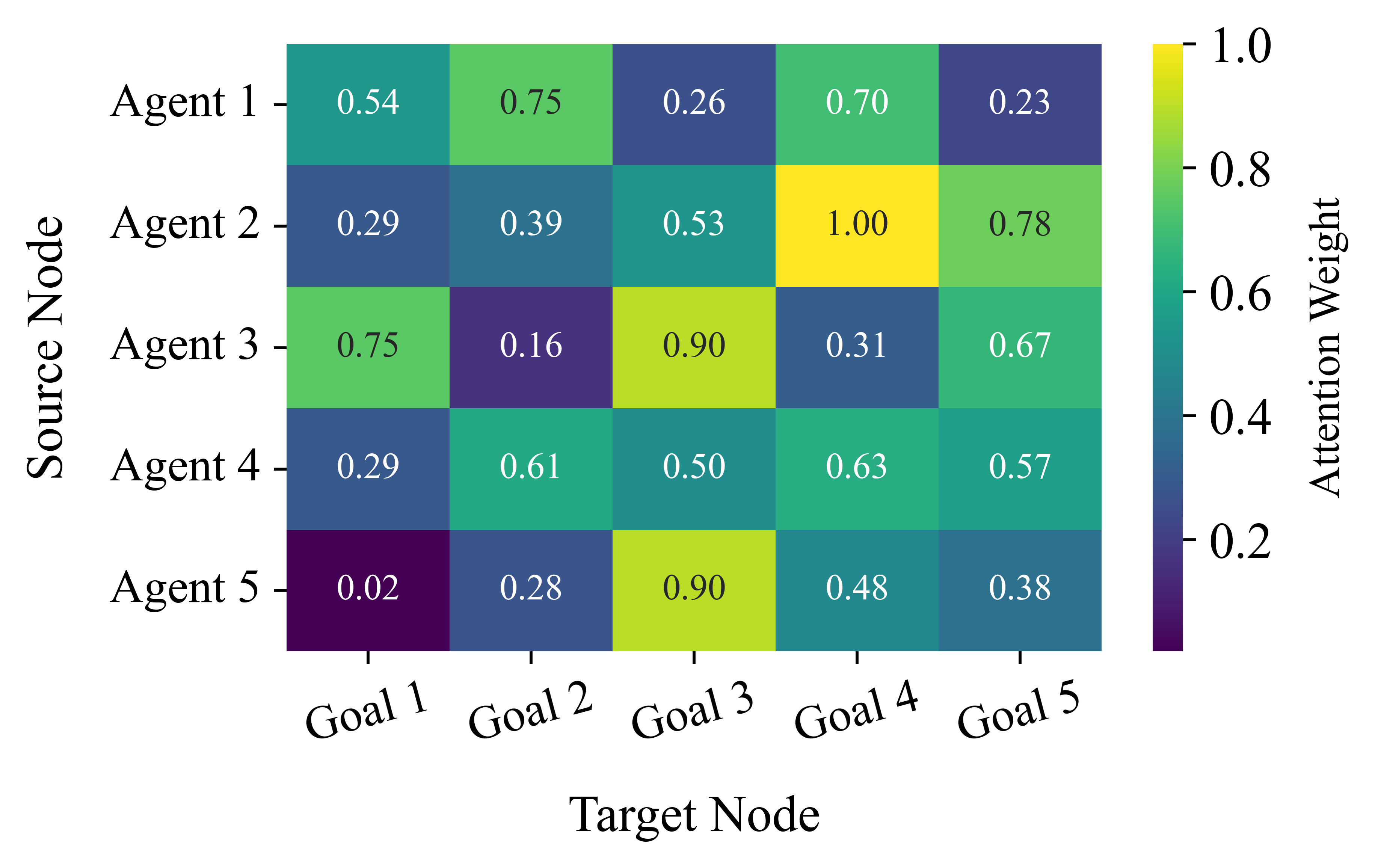}
    \caption{The heatmap of attention weights generated by the transformer-based architecture. Rows represent source nodes (agents), and columns represent target nodes (goals). Brighter colors (higher weights) indicate stronger attention, highlighting critical agent-goal relationships.}
    \label{fig:heatmap}
\end{figure}

The heatmap (Fig.~\ref{fig:heatmap}) illustrates the attention weights from the transformer-based architecture, where rows represent source nodes (agents), and columns represent target nodes (goals) in the constructed graph. Each cell indicates the weight assigned to the corresponding edge, reflecting the priority of the interaction between an agent and a goal. For example, the cell in row 2 and column 4 (Agent 2 and Goal 4) shows a weight of 1.00, indicating the strongest possible attention and highlighting that Agent 2 prioritizes Goal 4 over all others. 

\subsection{Learning Algorithm}

We employ a Double DQN framework with prioritized experience replay to optimize agent behavior. The Q-network minimizes the temporal difference error using the following loss function:

\begin{equation}
    \mathcal{L}(\theta) = \mathbb{E}_{(s,a,r,s')\sim\mathcal{D}}[w_i(y_i - Q(s,a;\theta))^2]
\end{equation}

\begin{equation}    
    y_i = r + \gamma Q'(s', \arg\max_{a'} Q(s',a';\theta);\theta').
\end{equation}

The importance sampling weights $w_i$ are computed as
\begin{equation}
    w_i = (\frac{1}{N} \cdot \frac{1}{(|\delta_i| + \epsilon)^\alpha})^\beta,
\end{equation}
where $N$ is the batch size, $\delta_i$ is the temporal difference error, and $\epsilon$, $\alpha$, and $\beta$ are hyperparameters.

\section{Experiment Results}


\subsection{Training Phase}
The network is trained using a designed set of parameters. Most notably, we use a learning rate of $\alpha=0.0005$, experience replay buffer size of 100,000, and a soft update rate of $\tau = 0.001$.
Each training update is performed with a mini-batch size of 64 experiences, selected using prioritized sampling to emphasize transitions with higher learning potential.
Exploration during training is guided by an $\epsilon$-greedy strategy, where the exploration rate $\epsilon$ decays linearly from 1.0 to 0.01 over the training process. 

Training proceeds in episodes of a maximum length of 200 steps, with the network updated every 4 steps using prioritized experience replay. The target network parameters $\theta'$ are updated using soft updates:
\begin{equation}
    \theta' \leftarrow \tau\theta + (1-\tau)\theta' ,
\end{equation}
where $\theta$ represents the online network parameters.


\subsection{Results and Analysis}

Table \ref{tab:configs} outlines the different environment configurations used in our experiments. For each configuration, we tested both the proposed GNN approach and a baseline DQN with training durations of 250,000 and 1,000,000 steps. Additionally, we compared the performance of our proposed method to well-respected methods, including Particle Swarm Optimization (PSO), Density-Based Scan (DBSCAN), Greedy Search, and conventional RL, all under arbitrarily-chosen, fully-equivalent configurations (5 agents, 76 goals in a 100×100 grid). 

\begin{table}[!h]
\caption{Environment Configurations}
\begin{center}
\begin{tabular}{|c|c|c|c|}
\hline
\textbf{Agents} & \textbf{Goals} & \textbf{Grid Size} & \textbf{Steps/Episode} \\
\hline
2 & 10 & 10×10 & 150 \\
4 & 20 & 20×20 & 150 \\
8 & 43 & 30×30 & 175 \\
15 & 76 & 40×40 & 200 \\
23 & 118 & 50×50 & 250 \\
33 & 169 & 60×60 & 300 \\
\hline
\end{tabular}
\label{tab:configs}
\end{center}
\end{table}

Performance analysis is performed using two key metrics:
\begin{itemize}
    \item Goal Collection Percentage: The proportion of goals successfully visited.
    \item Grid Coverage Percentage: The proportion of the environment grid observed by agents during an episode.
\end{itemize}
 
\begin{figure}[!h]
\centering
\includegraphics[width=0.95\columnwidth]{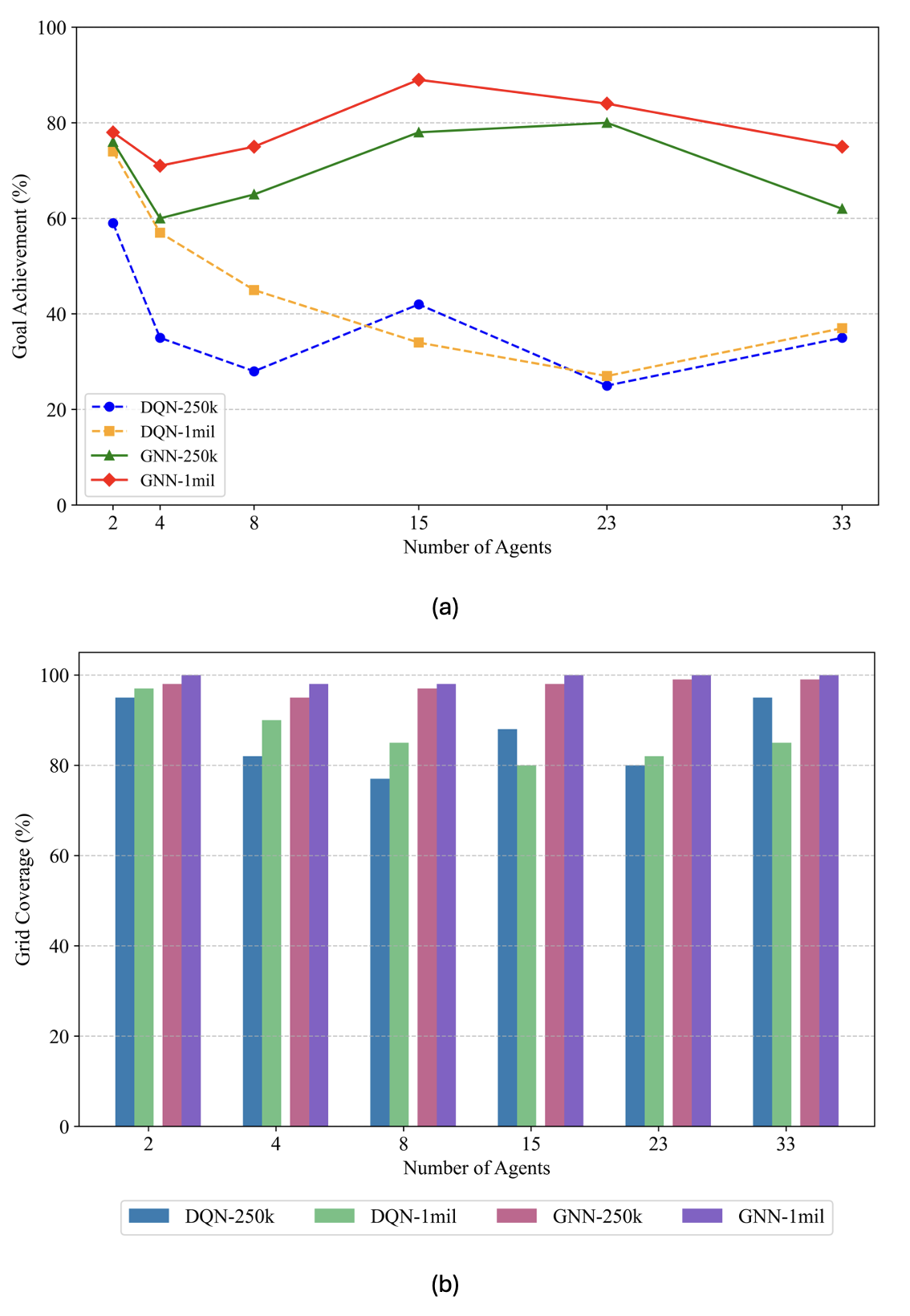}
\caption{Comparison between the proposed method and the baseline DQN method in terms of (a) goal achievement and (b) grid coverage.}
\label{fig:comparison}
\end{figure}

The results in Fig.\ref{fig:comparison}.a. demonstrate that the GNN architecture consistently outperforms the baseline DQN, with the performance gap widening as the environment scale increases. This suggests that the GNN's ability to process and leverage inter-agent relationships becomes increasingly valuable in larger, more complex scenarios. In particular, in a large grid environment with 15 agents, our proposed method achieves a 90\% goal collection rate, whereas the best-performing DQN configuration only reaches 42\%, marking a 110\% relative improvement in performance.
Additionally, the GNN achieved near-complete grid coverage in all configurations, whereas DQN struggled with larger environments, achieving only 82\% coverage in the largest configuration as shown in Fig.\ref{fig:comparison}.b.


\begin{figure}[!h]
\centering
\includegraphics[width=\columnwidth]{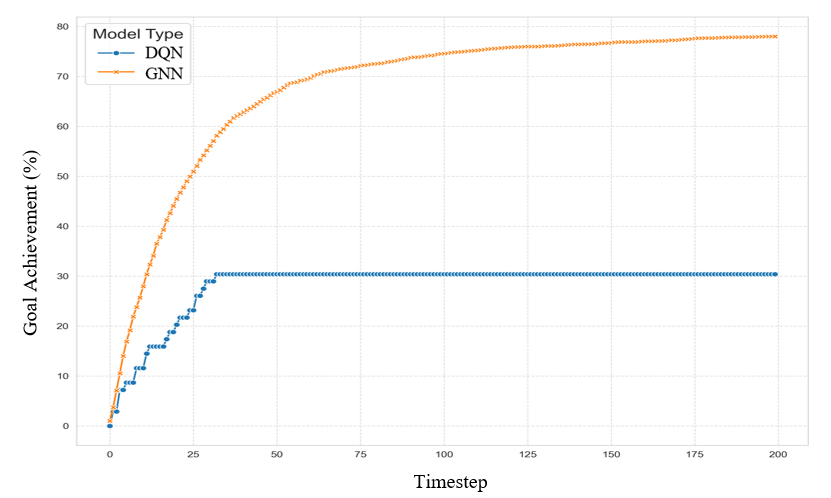}
\caption{Performance comparison between GNN and DQN approaches showing the percentage of goals collected over time with 15 agents. }
\label{fig:timestep}
\end{figure}

Further analysis of the temporal behavior shown in Fig. \ref{fig:timestep} reveals how each method performs in collecting goals over an episode. The GNN demonstrated higher efficiency, collecting goals more rapidly than DQN. 

\begin{figure}[!h]
    \centering
    \includegraphics[width=\columnwidth]{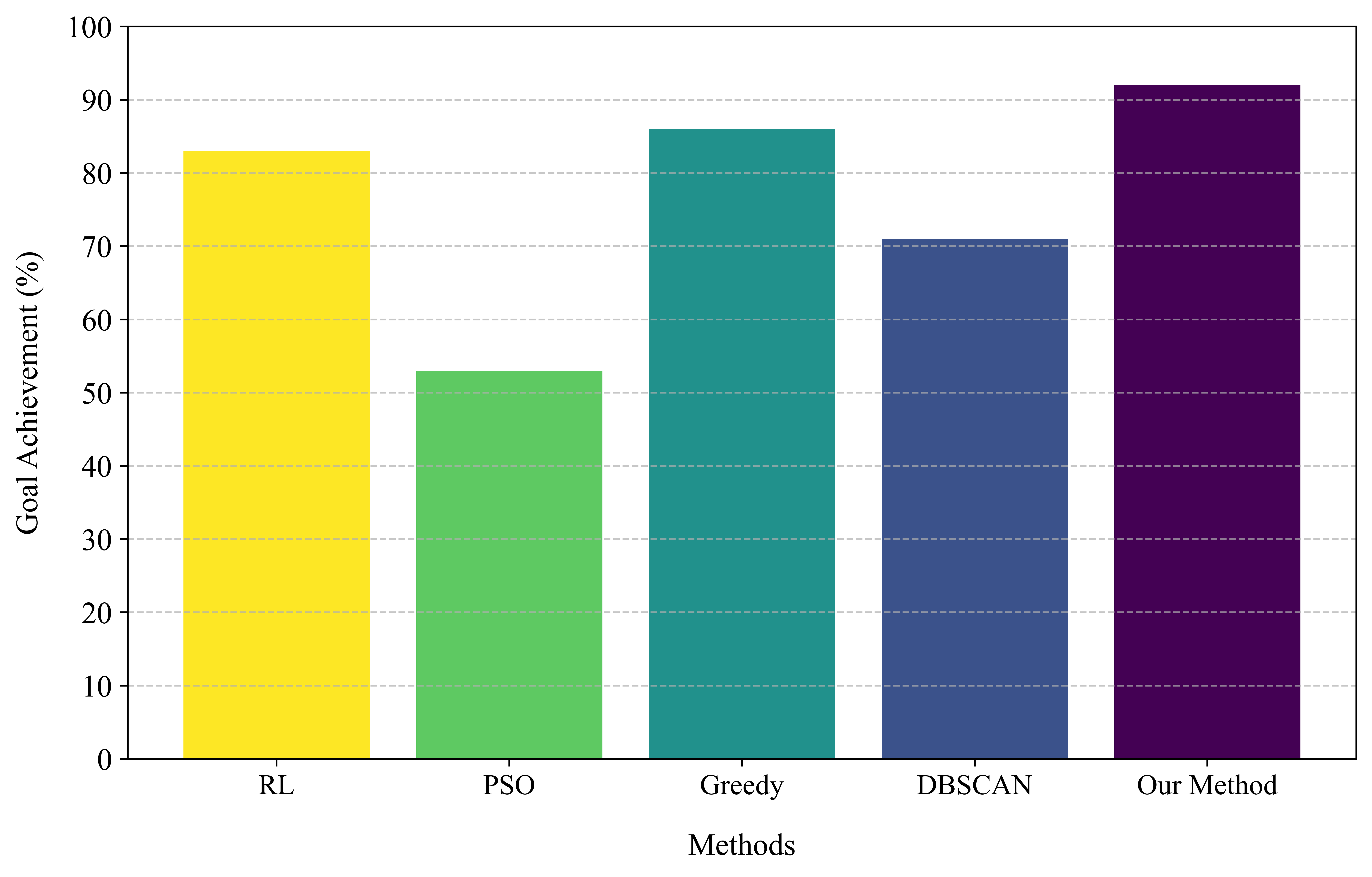}
    \caption{Comparison of goal achievement between the proposed method and benchmark algorithms.}
    \label{fig:benchmark}
\end{figure}

Fig.\ref{fig:benchmark} provides comparative results for the proposed GNN-based framework along with broadly used competitor algorithms, including PSO, DBSCAN, Greedy Search, and simple RL, in a 100×100 grid environment with 5 agents and 76 goals. The results highlight the superiority of the GNN approach. 
Another important metric, the number of steps per episode, underscores the efficiency of our proposed method. The GNN-based approach required an average of 200 steps to complete an episode, significantly outperforming the average of 600 steps taken by the other methods. These results emphasize the framework’s capability to efficiently coordinate agents and optimize task performance, particularly in dynamic and partially observable environments.

\begin{figure}[!h]
\centering
\includegraphics[width=\columnwidth, height=0.25\textheight]{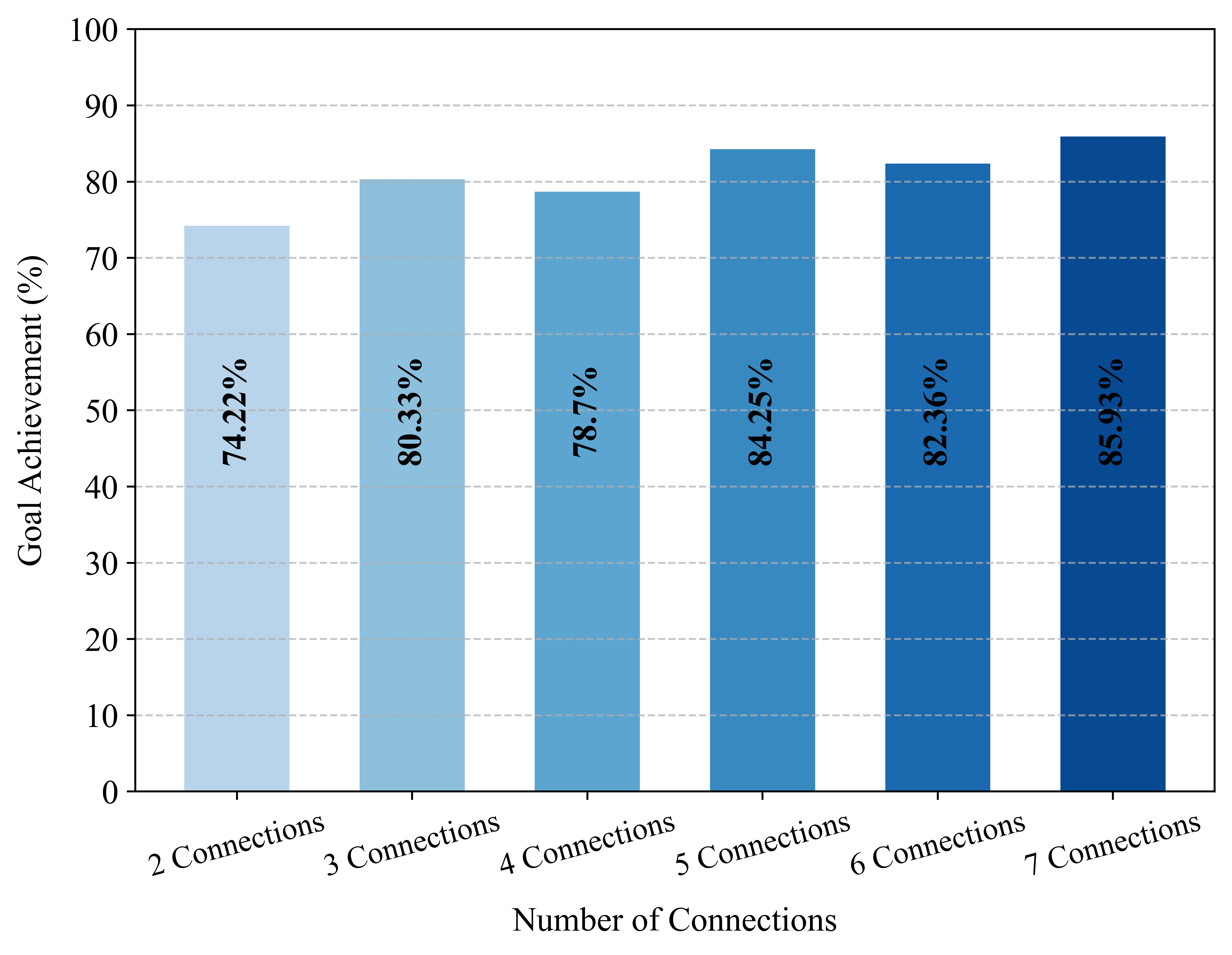}
\caption{Impact of varying agent connection limits on GNN performance in 40×40 environment with 15 agents. The analysis covers connection limits from 2 to 7 nearest neighbors.} 
\label{fig:connectivity}
\end{figure}

Furthermore, we conducted an ablation study varying the number of agent connections in the 40×40 environment configuration (15 agents, 76 goals) in Fig. \ref{fig:connectivity} to investigate the role of inter-agent communication in our framework. 
The results demonstrate that increasing the number of connections initially improves system performance, as agents benefit from enhanced information exchange and coordination. However, the improvement saturates beyond a certain threshold, suggesting diminishing returns as communication overhead increases. This observation highlights the trade-off between communication bandwidth and performance. This balance provides a practical guideline for designing communication-constrained multi-agent systems.

\section{Conclusion}

This paper presented a novel framework integrating Graph Neural Networks (GNNs), transformers, and Double Deep Q-Networks (DDQN) to enhance multi-agent navigation and task execution in dynamic environments. The proposed approach leverages GNNs for structured inter-agent communication, transformers for sequence-based decision-making, and DDQN for policy optimization. Experimental results demonstrate the superiority of our method over benchmark approaches, achieving 90\% goal collection and 100\% grid coverage while significantly reducing the number of steps per episode. These results highlight the effectiveness of our framework in enabling efficient, scalable, and cooperative UAV path planning.


\end{document}